\documentclass[twoside]{article}
\usepackage{graphicx}
\usepackage{fancyhdr}
\usepackage{multicol}
\usepackage{styl-ap}

\begin{document}
\title{The Impact of Suzaku Measurements on Astroparticle Physics}
\author{Naomi Ota\work{1}}
\workplace{Department of Physics, Nara Women's University, Kitauoyanishimachi, Nara, Nara 630-8506, Japan
}
\mainauthor{naomi@cc.nara-wu.ac.jp}
\maketitle

\begin{abstract}%
  Results from the {\it Suzaku} X-ray broad-band observations of
  clusters of galaxies are summarized. Aiming at understanding the
  physics of gas heating/particle acceleration and the cluster
  dynamical evolution, we search for non-thermal hard X-ray emission
  from merging clusters, particularly A2163 and the Bullet Cluster,
  based on the {\it Suzaku} and {\it XMM-Newton}/{\it Chandra} joint
  analyses. The observed hard X-ray emission is well represented by
  single- or multi-temperature thermal models, including super-hot
  ($kT\sim20$~keV) gas. However, no significant non-thermal
  hard X-ray emission has been detected. Together with the presently
  available literature, the hard X-ray properties have been studied
  for about 10 clusters with {\it Suzaku}. The present status on {\it
    Suzaku} measurements of non-thermal X-ray emission and the cluster
  magnetic field are summarized and compared with those from the {\it
    RXTE}, {\it BeppoSAX}, and {\it Swift} satellites. The future
  prospects are briefly mentioned.
\end{abstract}

\keywords{Clusters of galaxies - X-ray spectroscopy - Astroparticle physics - }

\begin{multicols}{2}
\section{Introduction}\label{sec:intro}
According to the standard scenario of the structure formation in the
Universe, clusters form via collisions and mergers of smaller groups
and clusters. A cluster merger has a kinetic energy of the order of
$10^{65}~{\rm erg}$.  This is the most energetic event in the Universe
since the Big Bang. If two such objects collide with each other, a
huge amount of energy may be released and a certain fraction is
expected to heat the gas and accelerate particles through shock waves,
and induce bulk and turbulent gas motions.

Signatures of cluster merging can be recognized in many ways. In the
X-ray band, irregular morphology and complex temperature structure of
the gas show that the system is disturbed due to past mergers.  At radio wavelengths, synchrotron emission extending over a Mpc scale
have been discovered from more than 30 clusters
\cite{giovannini99}. The existence of radio halo emission suggests
that relativistic electrons are being accelerated in the intracluster
space. Interestingly, there is a correlation between the non-thermal,
radio synchrotron power ($P_{1.4}$) and thermal X-ray luminosity
($L_X$), for merging clusters, while relaxed clusters without a radio halo
lie in a region well separated from the merging clusters on the
$P_{1.4}-L_X$ plane \cite{brunetti09}. It is suggested that generation
of high-energy particles is connected to the dynamical evolution of
clusters \cite{cassano10}. From the observational point of view,
however, a direct link between the radio synchrotron halo and the
non-thermal hard X-ray property is yet to be clarified.

In X-rays, the same population of high-energy electrons is thought to
interact with 3K CMB photons and then generate non-thermal
Inverse-Compton (IC) emission. The IC emission in excess of the
thermal emission is then predicted to be seen in the hard X-ray band
(typically, above 10~keV) where the thermal emission diminishes.  In
addition, from radio observation alone, we cannot separate the
energy of magnetic fields and the energy of high-energy
electrons. However, by comparing the radio and hard X-ray fluxes, the
cluster magnetic field is also estimated under the assumption that
the same population of relativistic electrons scatters off of CMB photons
\cite{kaastra08}.  Models for non-thermal bremsstrahlung emission
caused by suprathermal electrons with energies of 10--200~keV are
presented in \cite{sarazin00}.

The existence of non-thermal IC hard X-ray emission has been reported
in nearby clusters. The archetype is the Coma cluster: the non-thermal
hard X-ray flux has been measured by {\it RXTE} \cite{rephaeli02} and by 
{\it BeppoSAX} \cite{fusco-femiano04}. Recent reports based on broad-band X-ray observations with {\it Suzaku} \cite{wik09} and {\it
  Swift} \cite{wik11} did not detect any significant non-thermal
component and the hard X-ray flux is reproduced by thermal models. The
mismatch among several satellites is suggested to be reconciled if
different sizes of fields-of-view are taken into
consideration\cite{fusco-femiano11}.

Difficulty in firm detection of non-thermal hard X-ray emission from
clusters seems to indicate that the IC emission is in fact very faint
and/or that its spatial distribution makes it difficult to detect 
using the present instruments with limited imaging capability. Another
issue is spectral modeling of the thermal emission component. At energies
$>10$~keV, the X-ray emission from hot clusters is still dominated by
thermal bremsstrahlung emission \cite{wik09,ota08}. Thus detailed
characterization of the thermal emission spectrum, including the
multi-temperature components, is indispensable for separating the
non-thermal emission from the thermal emission.  As is discovered in
RX~J1347.5--1145 \cite{ota08}, the hard X-ray emission in the {\it
  Suzaku}/Hard X-ray Detector (HXD) band is originated predominately
by the super-hot ($kT\sim 25$~keV) gas in the cluster. Such violent
mergers often produce shock-heated gas with high ($\gg 10$~keV)
temperature. This needs to be properly taken into account in the
thermal modeling in order to study high-energy populations in the
intracluster space.

\section{Objectives}
We search for non-thermal hard X-ray emission from merging clusters
using {\it Suzaku} broad-band X-ray spectroscopy to reveal the
origin of hard X-ray emission and obtain new insight into the gas
physics and the cluster dynamical evolution. We also estimate the
magnetic fields in clusters by comparing the hard X-ray and
radio fluxes. Finally, we compare the {\it Suzaku} results with those
from other satellites and summarize the present status of hard X-ray
studies of clusters and briefly discuss the future prospects.

We adopt a cosmological model with $\Omega_{\rm M}=0.27$,
$\Omega_\Lambda=0.73$, and the Hubble constant $H_0=70~{\rm
  km\,s^{-1}\,Mpc^{-1}}$ throughout the paper.  Quoted errors indicate
the 90\% confidence intervals, unless otherwise specified.

\section{Sample and the {\it Suzaku} data}
The 5th Japanese satellite, {\it Suzaku} \cite{mitsuda07}, is equipped
with X-ray CCD imaging spectrometers (XIS; \cite{koyama07}) and
HXD \cite{takahashi07} and enables sensitive broad-band X-ray
observations thanks to the low and stable background
levels. In particular, the PIN detectors of the HXD instrument are
useful for the present study because they have achieved a lower 
background level than other missions in the 10--60~keV range
\cite{fukazawa09}. In order to constrain the hard X-ray
properties of merging clusters, we conducted {\it Suzaku}
observations of several targets. We focus on X-ray bright sources
with radio synchrotron halos, particularly A2163 at $z=0.203$, and
the Bullet Cluster at $z=0.296$, since they are considered to have
undergone a recent, violent merger.  The {\it Suzaku}'s exposure times
are 113~ksec and 41~ksec for A2163 south and north-east regions, and
80~ksec for the Bullet Cluster.

Figure~1 shows the {\it XMM-Newton} image of A2163 with overlaid the
{\it Suzaku}/HXD field of views for the two pointing
observations. A2163 is the brightest Abell cluster hosting one of the
brightest radio halos \cite{govoni04,feretti04}. Previous X-ray
observations showed the presence of a high temperature region in the
north-east, which can be attributed to the merger shock
\cite{bourdin11,markevitch01}. In the hard X-ray energies,
\cite{rephaeli06} reported the detection of non-thermal IC emission
with a flux level of $F_{\rm NT} =
1.1^{+1.7}_{-0.9}\times10^{-11}~{\rm erg\,s^{-1}\,cm^{-2}}$ in the
20--80~keV band from the long {\it RXTE} observations, while
\cite{feretti01} derived an upper limit of $F_{\rm
  NT}<5.6\times10^{-12}~{\rm erg\,s^{-1}\,cm^{-2}}$ from the {\it
  BeppoSAX}/PDS observations. Thus the {\it Suzaku} observation will
further provide independent measurements of hard X-ray properties of
the sample.

\begin{myfigure}
\centerline{\resizebox{90mm}{!}{\includegraphics{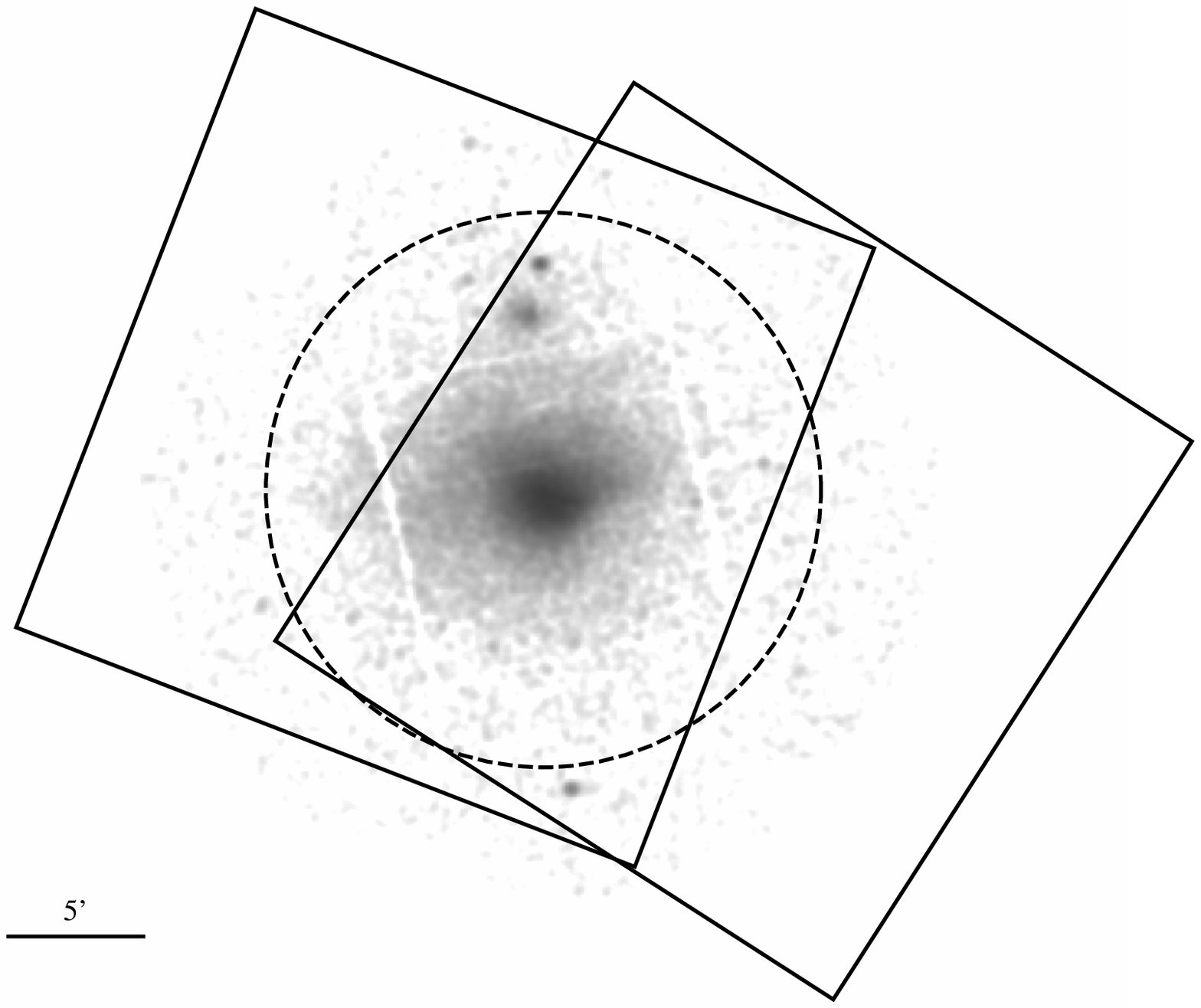}}}
\caption{{\it XMM-Newton} image of A2163 (grayscale) with overlaid the
  {\it Suzaku} HXD field of views ($34'\times 34'$ in FWHM) for the
  two pointing observations (the solid boxes). The {\it XMM-Newton}
  integration region for the global spectrum is indicated with the
  dashed circle.}\label{fig1}
\end{myfigure}

The Bullet Cluster exhibits extended radio halo emission
\cite{liang00} and a prominent strong shock feature
\cite{markevitch02}. The hottest gas with $kT>20$~keV exists in the
region of the radio halo enhancement. Possible detection of
non-thermal hard X-ray emission has been reported; the non-thermal
flux in the 20--100~keV band is $(0.5\pm 0.2)\times10^{-11}~{\rm
  erg\,s^{-1}\,cm^{-2}}$ by {\it RXTE} \cite{petrosian06} and
$3.4^{+1.1}_{-1.0}\times10^{-12}~{\rm erg\,s^{-1}\,cm^{-2}}$ by {\it
  Swift} \cite{ajello10}. Thus it is worth examining the existence of
IC emission with a detailed {\it Suzaku} analysis.

\section{Analysis strategy}
To accurately measure the non-thermal X-ray emission from clusters at
hard X-ray energies ($>10$~keV), we need:
\begin{enumerate}
\item a careful assessment of the background components, 
\item detailed modeling of the thermal emission.
\end{enumerate}
For 1., the {\it Suzaku} HXD background consists of the Cosmic X-ray
Background and the instrumental Non-X-ray Background (NXB), and is
dominated by the latter. We estimate the systematic error of NXB to be
2\% by analyzing the HXD data during the Earth-occultation and quote
the 2\% systematic error ($1\sigma$) in the spectral analysis of A2163
and the Bullet Cluster. This is consistent with SUZAKU-MEMO 2008-03 by
the instrument
team\footnote{http:\//\//www.astro.isas.jaxa.jp\//suzaku\//doc\//suzakumemo\//suzakumemo-2008-03.pdf}.

For 2., since the merging clusters can have a complex, multi-temperature
structure, including a very hot thermal gas that emits hard X-rays, we
apply single-, two-, and multi-temperature thermal models to the {\it
  Suzaku} spectra. As detailed later, the multi-temperature model is
constructed based on the analysis of {\it XMM-Newton} or {\it Chandra}
data. The {\it Suzaku} and {\it XMM-Newton}/{\it Chandra} joint
analysis allows us to take advantage of {\it Suzaku}'s spectral
sensitivity in the wide X-ray band and the {\it XMM-Newton}/{\it
  Chandra}'s high spatial resolution, as demonstrated by \cite{ota08}.

\section{Results}
\subsection{A2163}\label{subsec:a2163}
With {\it Suzaku}/HXD, we detected X-ray emission from the cluster up
to 50~keV. With the CXB and NXB components subtracted, the 12--60~keV
source flux is measured to be $1.52\pm0.06\, (\pm 0.28)\times
10^{-11}~{\rm erg\,s^{-1}\,cm^{-2}}$. Here the first and second errors
indicate the 1-$\sigma$ statistical and systematic uncertainties.  The
1-$\sigma$ systematic error of the flux is estimated by changing the
normalization of the NXB model by $\pm 2$\%.  Thus the detection of
hard X-ray emission is significant at the $>5\sigma$ level even if the
systematic error of NXB is considered.

Next we performed the joint {\it XMM}+{\it Suzaku}/HXD spectral analyses
under i) the single-temperature model, ii) the two-temperature model, and iii)
the multi-temperature (multi-T) model. For i) and ii), the {\it
  XMM-Newton} PN and MOS spectra were accumulated from the $r=10'$
circular region, as shown in Figure~1.

For i), the {\it XMM}+{\it Suzaku} broad-band spectra in the 0.3--60~keV
band can be fitted by the APEC thermal emission model: the resultant
gas temperature and the metal abundance are $kT=13.5\pm 0.5$~keV and
$Z=0.29\pm 0.10$~solar, respectively, and $\chi^2$/d.o.f. is
1249/1180. Thus the hard X-ray emission is likely to be dominated by
the thermal emission. For ii), we have not found any significant
improvement of the fits in comparison to case i). However, there is obviously a complex temperature structure as shown by
the previous X-ray observations, and we attempted to construct the multi-T
model for the thermal emission below.

iii) The $10'$ spectral region of the {\it XMM} data is divided into
$2'\times2'$ grids and the single-component APEC model is assumed for
each grid. Figure~2 shows the total model of multi-T components
(namely the sum of the single APEC models for the grids) fit to the
{\it Suzaku}/HXD data. We find that this multi-T model gives an
acceptable fit to the HXD spectrum. Note that the model includes a
$kT\sim18$~keV APEC component for the north-east ``shock'' region,
which is in agreement with the previous {\it XMM} result
\cite{bourdin11}. The X-ray luminosity of this hot component is
$5\times10^{44}\,{\rm erg\,s^{-1}}$, which contributes to the hard
X-ray emission by about 15\%.

\begin{myfigure}
\centerline{\resizebox{75mm}{!}{\includegraphics{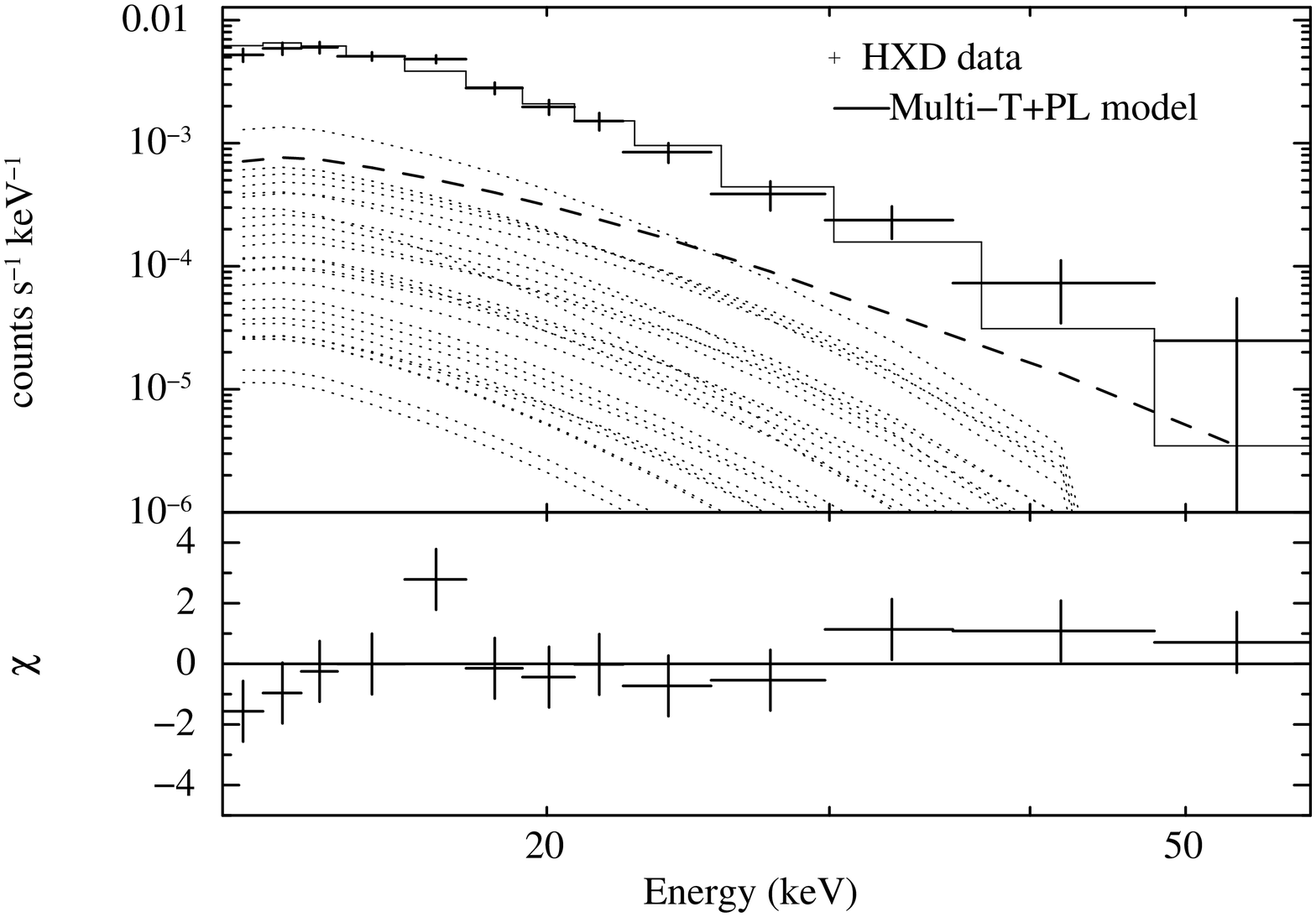}}}
\caption{Upper panel: the {\it Suzaku}/HXD spectrum of A2163 (the
  crosses) fitted with the multi-T + power-law model (the solid step
  function). The contribution from the spectral components, i.e.,
  multiple temperature components (the dotted lines) and the
  non-thermal power-law model (the dotted lines) are indicated in the
  figure. Bottom panel: the residual of the fitting is shown with the
  crosses.}\label{fig2}
\end{myfigure}

In the case of iii), the additional power-law component for
non-thermal IC emission does not significantly improve the fit to the
observed HXD spectrum. Assuming the multi-T+ power-law model with
$\Gamma=2.18$ (i.e., the same index in radio \cite{feretti04}), the
90\% upper limit on the non-thermal emission is derived as
$9.4\times10^{-12}~{\rm erg\,s^{-1}\,cm^{-2}}$.  This gives 3-times
stronger constraints than the upper limit obtained from {\it RXTE}
\cite{rephaeli06}, though the present result is consistent with {\it
  RXTE} and {\it BeppoSAX} \cite{feretti01} within their errors.

More details of data reduction and discussions are to be described in
a forthcoming paper (N. Ota et al. in preparation).

\subsection{The Bullet Cluster}
Following similar methods mentioned in \S\ref{subsec:a2163}, we
analyzed the {\it Suzaku} and {\it Chandra} data of the Bullet
Cluster.  Our preliminary analysis showed that the global cluster
spectrum taken with {\it Suzaku} is well represented by a
two-component thermal model. However, the non-thermal
emission is not significantly detected, yielding the 90\% upper limit
on a $\Gamma=1.5$ power-law component to be $\sim 10^{-11}~{\rm
  erg\,s^{-1}\,cm^{-2}}$ (20--100~keV). Note that the derived upper
limit depends on the assumed photon index.

The observed broad-band {\it Suzaku} spectrum in the 1--50~keV band
can also be fitted by the multi-T model. Here the multi-T model is
constructed by analyzing the deep {\it Chandra} observations. We did
not confirm the previous IC detections with {\it RXTE} and {\it Swift}, 
and the present results support the thermal origin of the hard X-ray
emission.  Under the multi-T model, the flux of the hot ($kT>13$~keV)
thermal component is estimated to be $F_{\rm hot}\sim
2\times10^{-12}~{\rm erg\,s^{-1}\,cm^{-2}}$ and comparable to the
non-thermal flux reported by \cite{ajello10}, while the IC flux is
suggested to be even lower than $F_{\rm hot}$. More details of the
analysis, including an assessment of systematic uncertainty due to the
{\it Suzaku}-{\it Chandra} cross-calibration, will be presented in
Nagayoshi et al. in preparation.

\section{Discussion}\label{sec:discussion}
\subsection{Non-thermal hard X-ray emission}
For the two hot clusters, A2163 and Bullet, the hard X-ray spectra
taken with {\it Suzaku} are likely to be dominated by thermal emission,
giving stronger limits on the non-thermal IC emission. We also note
that a super-hot ($\sim 20$~keV) gas exists in both clusters whose
contribution is not negligible in the hard X-ray band. Thus this
component should be properly modeled in order to accurately measure
the non-thermal component. Furthermore, this hot gas is over-pressured
and thus thought to be short-lived, $\sim 0.5$~Gyr
\cite{ota08}. Therefore the existence of super-hot gas supports the
scenario of a recent merger.

Given the link between the radio flux and X-ray luminosity (see
\S\ref{sec:intro}), the generation of high-energy particles in
intracluster space is likely to be connected with the merging
events. Thus, {\it is there any relationship between $L_X$ and IC flux
  ($F_{\rm NT}$) or between the gas temperature ($kT$) and $F_{\rm NT}$?}

With {\it Suzaku}, non-thermal hard X-ray emission has been
constrained in about ten clusters. In Figure~3, $F_{\rm NT}$ is
plotted as a function of $kT$. The {\it Suzaku} results for nearby
clusters as well as RX~J1347.5--1145 were taken from the literature
\cite{kitaguchi07,kawaharada10,kawano09,wik09,sugawara09,nakazawa09,
  fujita08, ota08}.  The {\it RXTE}/{\it BeppoSAX}/{\it Swift} results
\cite{kaastra08,ajello09,ajello10,wik11} are also quoted for
comparison.  Note that the non-thermal flux depends on the modeling of the 
thermal component as well as the assumption of the photon index for the 
non-thermal component. For {\it Suzaku}, non-thermal emission has not
been detected and the upper limit was derived for the nearby objects
including the Coma cluster, A2319 etc. Hence there is no clear
relation on the $F_{\rm NT}- T$ plane.

\subsection{Cluster magnetic field}
Using the observed radio flux, $S_{\rm syn}$, and the relation $S_{\rm
  IC}/S_{\rm syn} = U_{\rm CMB}/U_{\rm B}$ \cite{kaastra08,ota12}, we
can infer the strength of the cluster magnetic field.  In the case of
A2163, adopting the radio flux $S_{\rm syn} = 155$~mJy at 1.4~GHz
\cite{feretti04} and the IC upper limit of $S_{\rm IC}<0.26~{\rm \mu
  Jy}$ at 12~keV from the this work, we obtain $B >0.09~{\rm \mu G}$
for the multi-T + power-law ($\Gamma=2.18$) model. For the Bullet
Cluster, the magnetic field is estimated to be $B > 0.06~{\rm \mu G}$
under the two-temperature APEC + power-law ($\Gamma=1.5$) model. Thus
a lower limit of the order of $\sim 0.1-1{\rm \mu G}$ has been
obtained in most of the clusters observed with {\it Suzaku}.
   

\begin{myfigure}
\centerline{\resizebox{75mm}{!}{\includegraphics{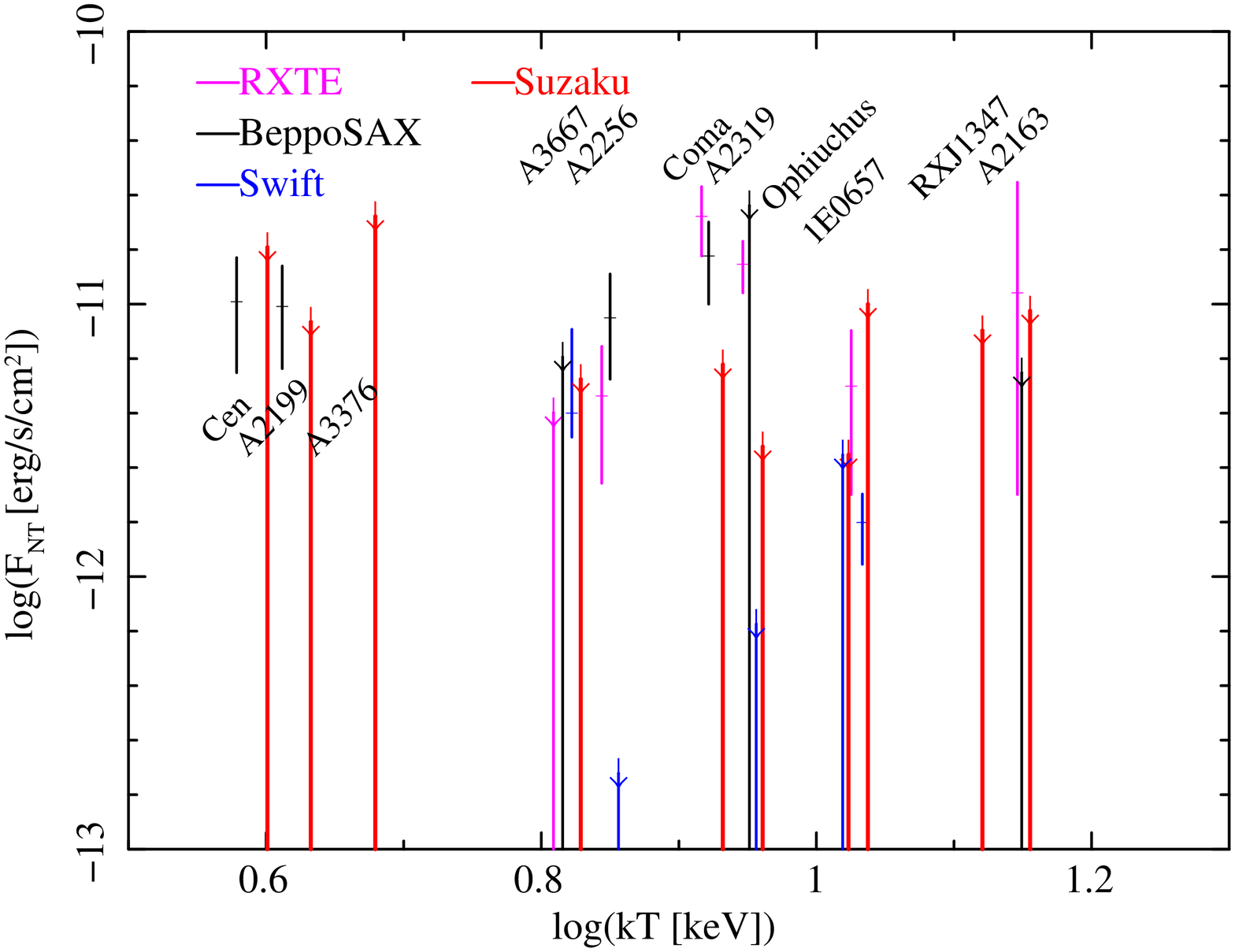}}}
\caption{Non-thermal IC hard X-ray flux in ten clusters measured by
  {\it Suzaku}. The results from {\it RXTE}, {\it BeppoSAX}, and {\it
    Swift} are also shown for comparison. See \S\ref{sec:discussion}
  for references.}\label{fig3}
\end{myfigure}

\section{Summary}
{\it Suzaku} broad-band X-ray observations of hot
clusters hosting a bright radio synchrotron halo, A2163 and the Bullet
cluster, were performed in order to search for the non-thermal hard X-ray component and reveal
the origin of the high-energy emission and the cluster dynamical
evolution. The {\it Suzaku}/HXD spectra for these
two clusters are well represented by single- or two-temperature
thermal models. Since the merging clusters can often have
shock-heated, super-hot ($kT\sim 20$~keV) gas, and its contribution is
not negligible in the interpretation of the hard X-ray origin, it is essential to determine the thermal component with high accuracy. This motivated us to carry out the joint {\it
  Suzaku}+{\it XMM-Newton}/{\it Chandra} spectral analysis under the
multi-T model. As a result, the observed {\it Suzaku} spectra are
consistent with the multi-T models and we did not find any significant
non-thermal IC emission from A2163 and the Bullet cluster.

The properties of hard X-rays and the cluster magnetic fields have
been studied for about ten clusters so far. There is no significant
detection of IC emission by {\it Suzaku}, giving the upper limits on the
level of $\sim10^{-11}~{\rm erg\,s^{-1}\,cm^{-2}}$.  Thus no clear
relationship between the non-thermal X-ray and radio properties or the
IC luminosity ($L_{\rm IC}$)--$T$ is seen.  To further constrain the
physics of gas heating and particle acceleration in clusters, it is
important to increase the number of cluster samples by applying the
present analysis method to other clusters.

Given that the relativistic particles are localized in small, shock regions,
present instruments with limited imaging capability may fail to
identify their existence. If this is the case, the advent of hard X-ray
imagers will benefit the cluster study.  Now {\it NuSTAR}
\cite{harrison10} is successfully in orbit and taking focused images
of the high-energy X-ray sky.  The {\it ASTRO-H} satellite is
scheduled to be launched in 2014 \cite{takahashi12}, and will carry
state-of-the-art instruments to realize sensitive X-ray observations
in the 0.3--600~keV band.  The Hard X-ray Imagers on {\it ASTRO-H}
will have an effective area comparable to {\it NuSTAR}.  These new
hard X-ray imagers will enable a more accurate high-temperature thermal
component and identification of the particle acceleration site to get
higher signal-to-noise ratios. This will lead us to detect the IC
emission to a level $\sim 2$ orders of magnitudes lower than the
present limit.  In addition, the high-spectral resolution of X-ray
micro-calorimeters will enable direct measurements of bulk/turbulent
gas motions\cite{mitsuda10}. {\it ASTRO-H} can therefore measure non-thermal energies in the form of kinetic gas motions (turbulence,
bulk gas motion) and relativistic particles, leading us to draw a more
comprehensive picture of the cluster structure and evolution.

\thanks N.O. thanks the organizers for the opportunity
to present this paper and R. Fusco-Femiano for useful suggestions and
discussions.

%

\end{multicols}
\end{document}